\documentclass[aps,prl,twocolumn,letterpaper,superscriptaddress]{revtex4-1}
\usepackage{amsmath}
\usepackage{amssymb}
\usepackage{amsfonts}
\usepackage{txfonts}
\usepackage{bbold}
\usepackage{color}
\usepackage{bm}
\usepackage{graphicx}
\usepackage{enumerate}
\usepackage{afterpage}

\newcommand{\brr}{{\mathbf r}}
\newcommand{\ecut}{\epsilon_{\rm{cut}}}

\begin{document}
\title{Characteristics of Two-Dimensional Quantum Turbulence in a Compressible Superfluid} 

\author{T. W. Neely}
\address{College of Optical Sciences, University of Arizona, Tucson, AZ 85721, USA}
\author{A. S. Bradley}
\address{Jack Dodd Centre for Quantum Technology, Department of Physics, University of Otago, Dunedin 9016, New Zealand}
\author{E. C. Samson}
\address{College of Optical Sciences, University of Arizona, Tucson, AZ 85721, USA}
\author{S. J. Rooney}
\address{Jack Dodd Centre for Quantum Technology, Department of Physics, University of Otago, Dunedin 9016, New Zealand}
\author{E. M. Wright}
\address{College of Optical Sciences, University of Arizona, Tucson, AZ 85721, USA}
\author{K. J. H. Law}
\address{Mathematics Institute, University of Warwick, Coventry CV4 7AL, UK}
\author{R. Carretero-Gonz\'{a}lez}
\address{Department of Mathematics and Statistics, San Diego State University, San Diego, CA 92182, USA}
\author{P. G. Kevrekidis}
\address{Department of Mathematics and Statistics, University of Massachusetts, Amherst, MA 01003, USA}
\author{M. J. Davis}
\address{School of Mathematics and Physics, University of Queensland, Qld 4072, Australia}
\author{B. P. Anderson\email{bpa@optics.arizona.edu}}
\address{College of Optical Sciences, University of Arizona, Tucson, AZ 85721, USA}

%\pacs{03.75.Nt,67.85.De,67.25.dk}

\begin{abstract}

Under suitable forcing a fluid exhibits turbulence, with characteristics strongly affected by the fluid's confining geometry.  Here we study two-dimensional quantum turbulence in a highly oblate Bose-Einstein condensate in an annular trap. As a compressible quantum fluid, this system affords a rich phenomenology, allowing coupling between vortex and acoustic energy. Small-scale stirring generates an experimentally observed disordered vortex distribution that evolves into large-scale flow in the form of a persistent current.  Numerical simulation of the experiment reveals additional characteristics of two-dimensional quantum turbulence: spontaneous clustering of same-circulation vortices, and an incompressible energy spectrum with $k^{-5/3}$ dependence for low wavenumbers $k$ and $k^{-3}$ dependence for high $k$.
\end{abstract}
\maketitle

A critical distinction between hydrodynamic turbulence in a bulk fluid \cite{Les2008.Turbulence} and in one whose flows are restricted to two dimensions is that energy dissipation at small length scales is generally inhibited in the latter.  In two-dimensional (2D) flows subject to small-scale forcing, energy flux is blocked through the small length scales and, instead, energy is transferred towards larger scales, comprising the inverse energy cascade of 2D turbulence \cite{2DT,2DCTreviews}.  Small-scale forcing may thus generate large-scale flows, as seen for instance in dispersal of atmospheric and oceanic particulates \cite{Kra2008.PF20.056602}, flows of soap films \cite{Mar1998.PRL80.3964,Riv2003.PRL90.104502} and plasmas \cite{Sha2005.PRE71.046409}, and Jupiter's Great Red Spot \cite{Som1988.Nat331.689,Mil1992.PRA45.2328}. However, the nature of 2D turbulence in \emph{quantum} fluids is less clear.  Progress in 2D quantum turbulence (2DQT) may offer innovative routes to understanding quantum fluid dynamics~\cite{Bar2001.book.Superfluid,Vin2002.JLTP128.167} and aspects of the universality of 2D turbulence.  Here we describe an experimental and numerical study of forced and decaying 2DQT in a dilute-gas Bose-Einstein condensate (BEC).  Our primary result is the first clear evidence that three key characteristics of 2D turbulence may also simultaneously appear in systems exhibiting 2DQT:  (i) emergence of large-scale flow from small-scale forcing, seen experimentally and numerically, (ii) numerical observation of the formation of coherent vortex structures accompanying approximate enstrophy conservation~\cite{enstrophy}, and (iii) numerical observation of an incompressible kinetic energy spectrum with $k^{-5/3}$ dependence for low wavenumbers $k$ and $k^{-3}$ dependence for high $k$.  Our observations are consistent with the notion that an inverse energy cascade can exist in this system.

Concepts of significance for 2D turbulence and quantum fluids share a common origin.   Analyzing point vortex motion in a bounded domain, Onsager proposed in 1949 that long-lived vortices may develop via mergers of smaller vortices in turbulent flows of a 2D fluid, enabling the remaining vortices to move more freely and thereby maximize system entropy \cite{Ons1949.NC6s2.279,Eyi2006.RMP78.87}.  He also proposed that vortices in superfluids have quantized circulation, and implied that turbulent 2D vortex dynamics might be ideally studied in superfluids.  However, experimental demonstration of 2DQT has not been reported and has only recently been addressed numerically~\cite{Naz2006.JLTP146.31,Hor2009.PRA80.023618,Num2010.JLTP158.415,Num2010.PRA81.063630,Gasenzer}.

To experimentally reach the 2DQT regime, we utilize optical and magnetic confinement to create highly oblate BECs  \cite{Nee2010.PRL104.160401}.    A harmonic potential with radial ($r$) and axial ($z$) trapping frequencies $(\omega_r/2\pi,\omega_z/2\pi) = (8,90)$~Hz confines BECs of up to $\sim$2$\times10^6$ $^{87}$Rb atoms having radial and axial radii of $52 \,\mu$m  and $5\, \mu$m respectively.  For these conditions, vortex bending and tilting are suppressed  \cite{Roo.PRA84.023637}, leading to 2D vortex dynamics.  An annular trap is created with a $23$-$\mu$m radius, blue-detuned Gaussian laser beam directed axially through the trap center; the beam creates a barrier of height $U_0\sim1.5 \mu_0$, where $\mu_0~\sim 8 \hbar \omega_z$ is the BEC chemical potential  in the purely harmonic trap.  Relative to the phase transition temperature $T_c \sim 116$ nK, the initial temperature is $T \sim 0.9 T_c$. 

%============================================================================
\begin{figure}
\begin{center}
\includegraphics[width=.9\columnwidth]{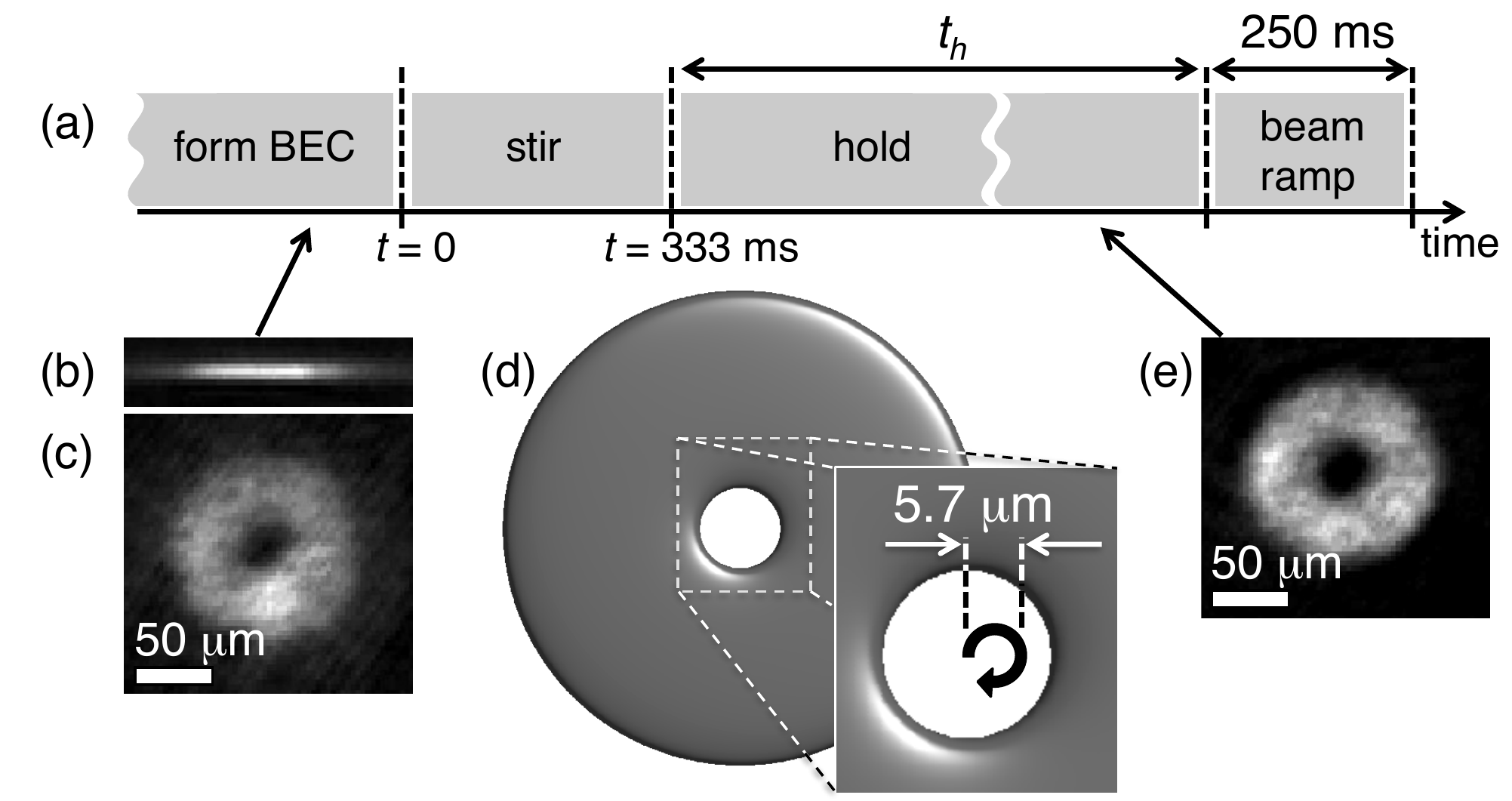}
\vspace{-3mm}
\caption{\label{fig1}
(a)  Timing sequence used to study 2DQT.  (b) and (c)  Experimental \emph{in situ} column-density images of the BEC immediately prior to the stir, as viewed (b) in the plane of 2D trapping and (c) along the $z$ axis. Lighter grayscale shades indicate larger column densities, as in subsequent experimental and numerical density data.  (d) Illustration of stirring, the black arrow shows the trajectory of the harmonic trap center relative to the larger fluid-free region created by the laser barrier.  (e) \emph{In situ} image of the BEC 10 s after stirring; vortices are not observable, necessitating an expansion stage to resolve them.}
\end{center}
\vspace{-5mm}
\end{figure}
%============================================================================

At time $t=0$ of each experimental run, a magnetic bias field moves the center of the harmonic trap, but not the central barrier, in one complete 5.7-$\mu$m-diameter circle over 333 ms. This motion induces small-scale forcing and nucleation of numerous vortices in a highly disordered distribution, which we identify with 2DQT much as the notion of a `vortex tangle' is identified with 3D quantum turbulence \cite{Bar2001.book.Superfluid,KetterleQT,Hen2009.PRL103.045301}.  Afterwards, the BEC remains in the annular trap for a variable hold time $t_h$ up to 50~s while the 2DQT decays. At $t_h=$1.17~s, the system temperature is reduced to $\sim$0.6$T_c$ in order to decrease rates of thermal damping and vortex-antivortex recombination.    At the end of the hold period, the central barrier is ramped off over 250 ms, the BEC is released from the trap, and ballistic expansion of the BEC enlarges the vortex cores such that they are resolvable by absorption imaging.  Figure~\ref{fig1} illustrates this sequence and shows images of the trapped BEC. 

\begin{figure}[t]
\begin{center}
\includegraphics[width=.95\columnwidth]{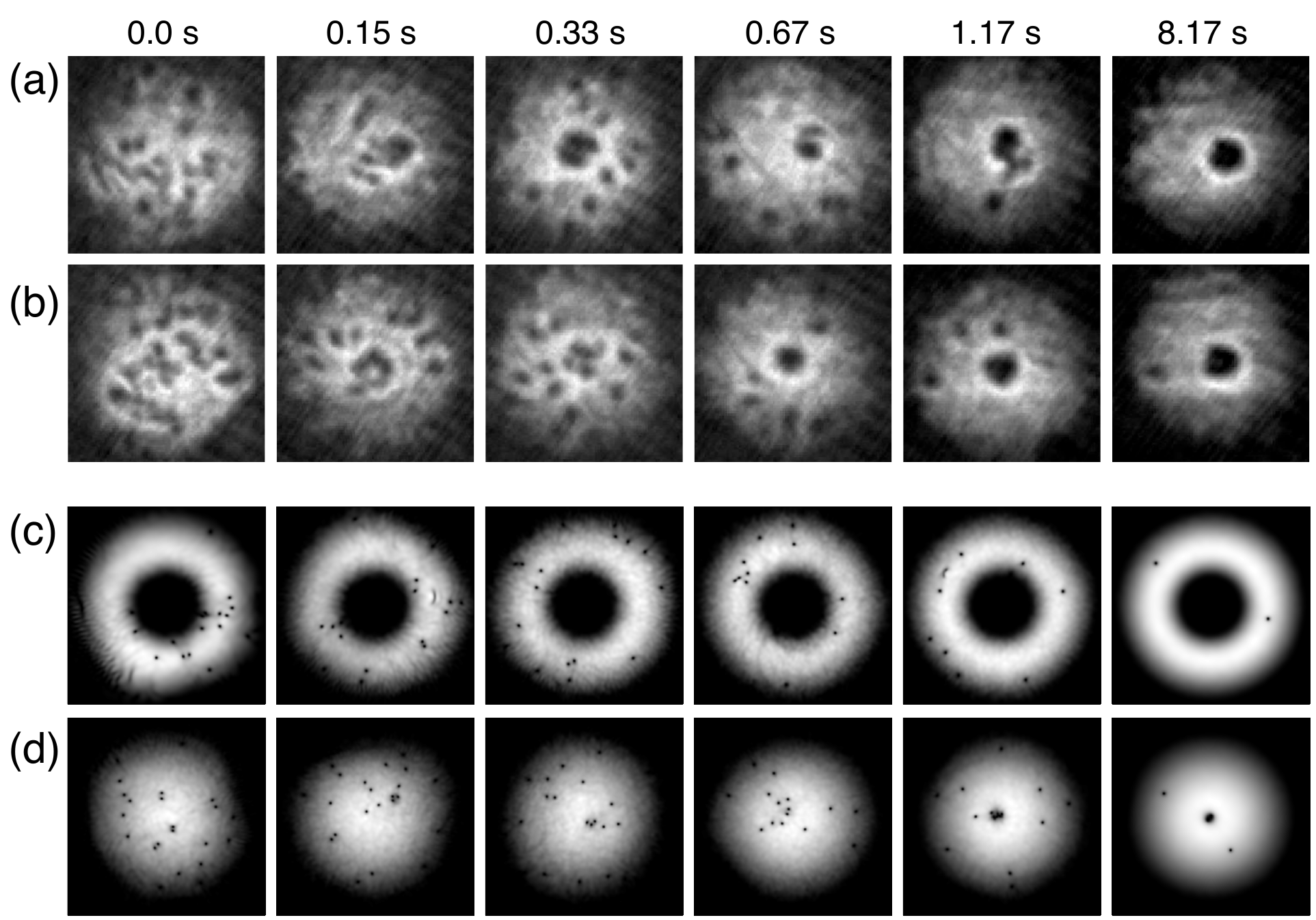}
\caption{\label{fig2}
(a) and (b) 200-$\mu$m-square experimental column-density images acquired at the hold times $t_h$ indicated above the images. Each BEC undergoes ballistic expansion immediately after the central barrier ramp-down in order to resolve the vortex cores. Each image is acquired from a separate experimental run.  (c)  \emph{In situ} numerical data (96-$\mu$m-square images) for the hold times indicated.  See also Movie S1 \cite{SM}. For each state represented in (c), ramping off the laser barrier in 250ms gives the data shown in (d).}
\end{center}
\vspace{-7mm}
\end{figure}

Two experimental time sequences of post-stir dynamics are shown in Figure~\ref{fig2}(a) and (b), emphasizing the microscopic variability of vortex distributions. Just after the stir ($t_h$ = 0 ms) a disordered vortex distribution appears. Large-scale superflow is evident after $t_h \approx 0.33$ s and with increasing $t_h$, as indicated by the large fluid-free hole in the middle of the expanded BEC; this flow evolves into a persistent current by $t_h \approx 8.17$~s.  An optional 3-s hold between barrier ramp-down and BEC release gives the vortices pinned by the central barrier time to separate enough to determine the circulation winding number about the barrier; see Supplemental Material Fig.~S1~ \cite{SM}. Our experiment demonstrates that under suitable conditions of forcing and dissipation, a highly disordered vortex distribution can evolve into a large scale flow in an annular trap.  However, measuring kinetic energy spectra and \emph{in situ} vortex dynamics remain forefront experimental problems, motivating us to utilize numerical modeling and analysis for further characterizing 2DQT in a stirred, trapped BEC.

BECs admit a first-principles theoretical approach that is numerically tractable, enabling accurate modeling~\cite{Bla2008.AIP57.363}.  The physical system consists of a large non-condensate component close to thermal equilibrium and a BEC responding both to external forcing and to damping by the non-condensate component. Numerically, we focus on the dynamics of just the BEC. 
We simulate the experimental stirring procedure using damped Gross-Pitaevskii theory~\cite{Pit1958}.  The parameters most readily measured are the total atom number $N$ and temperature $T$. We have developed an efficient Hartree-Fock scheme for determining the chemical potential $\mu(N,T)$ and reservoir cutoff energy $\epsilon_{\rm cut}(N,T)$ in Ref.~\cite{Roo.PRA81.023630}, and adapt the same procedure to the present experiment, accounting for the shift in the trap minimum caused by the central barrier.   We thus find the parameters needed to model the experimental system \cite{SM}.

Figure~\ref{fig2}(c) shows simulations that correspond to experimental observations. Here too vortices become pinned to the central barrier to form a persistent current; at  $t_h$=8.17 s, three vortices are pinned to the barrier, as indicated by the corresponding quantum phase profile (see Movie S1 \cite{SM}). Ramping off the obstacle beam in the simulation (over 250 ms as in the experiment) gives the column densities shown in Figure~\ref{fig2}(d), with density distributions more readily compared to (a) and (b). Development of superflow at $t_h=8.17$ s in (c) leads to a large region of low density in the trap center after barrier ramp-down, as seen in  (a), (b), (d).

%============================================================================
%============================================================================

Analysis of our numerical simulations further characterizes 2DQT through two distinct dynamical features of the system evolution, namely the development of a logarithmically bilinear kinetic energy spectrum, and the formation of tightly bound, long-lived clusters of vortices with the same sign of circulation.  To examine numerically the dependence of the kinetic energy on the wavenumber $k$ at any instant in time, we use the techniques of previous studies \cite{Hor2009.PRA80.023618,Num2010.JLTP158.415,Num2010.PRA81.063630,Gasenzer,Nor1997.PF9.2644} for extracting $E^i(k)$, the portion of a BEC's kinetic energy spectrum that corresponds to an incompressible superfluid component, derived by extracting the divergence-free density-weighted velocity field that embeds vorticity; the curl-free part of this field corresponds to sound waves and acoustic energy.  

\begin{figure}[t]
\includegraphics[width=.85\columnwidth]{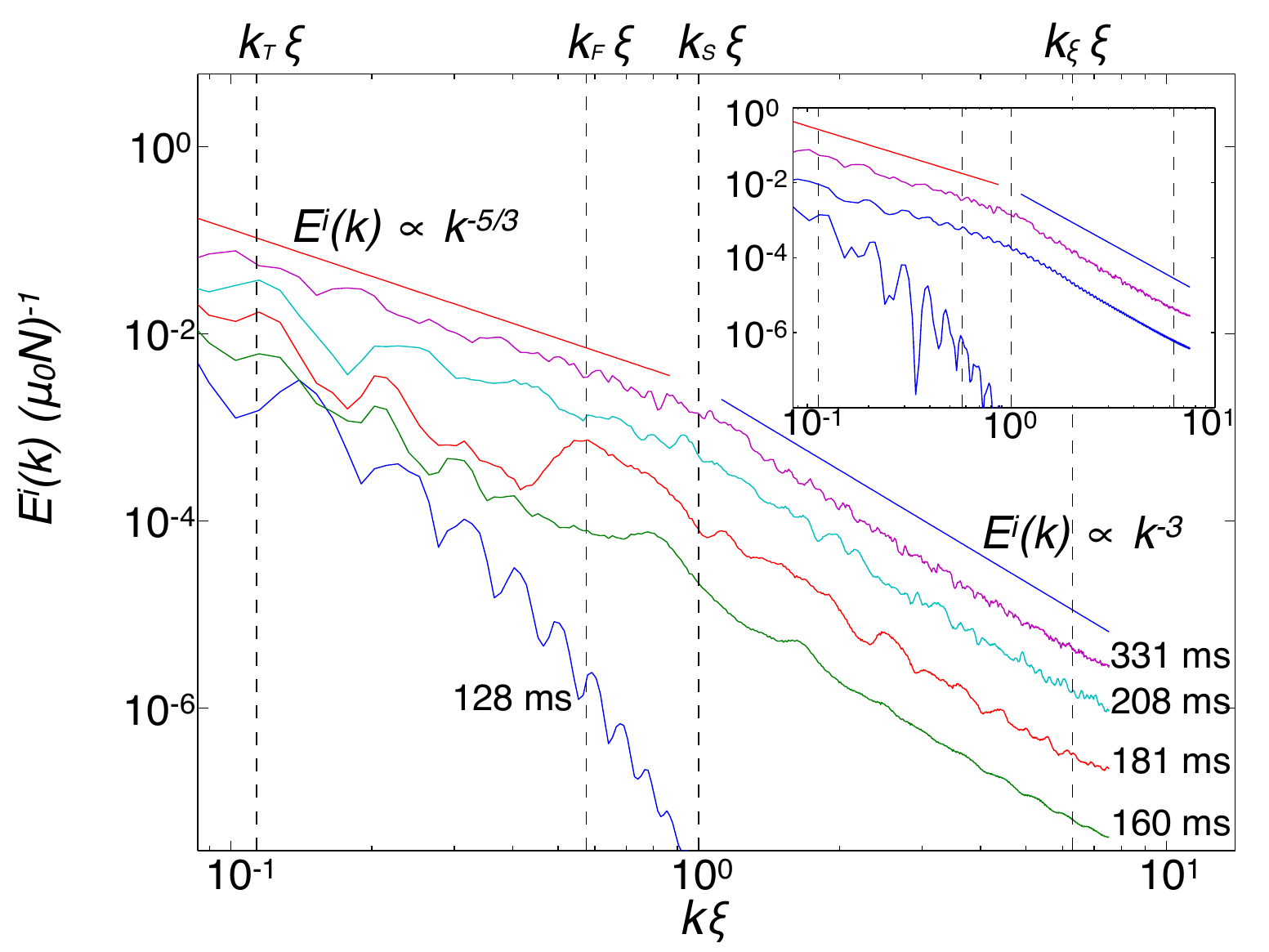}
\vspace{-3mm}
\caption{\label{fig3}
Log-log plots of $E^i(k)$ (per atom) for times $(128, 160, 181, 208, 331)$ ms over which forcing occurs, plotted against $k\xi$ with healing length $\xi =0.42\, \mu$m.    Vertical dashed lines indicate $k_T$, $k_F$, $k_s$, and $k_{\xi}$, defined in the text.  Red and blue lines indicate $E^i(k) \propto k^{-5/3}$ and $k^{-3}$, respectively.  A spectral peak at $181$ ms appears at $k_F \approx 2\pi/(11\xi)$. Inset:  log-log plot of $E^i(k)$ vs.~$k\xi$ (labels omitted) over the same domain as the main plot.  From top to bottom, curves show $E^i(k) \propto k^{-5/3}$ and $k^{-3}$ (solid lines), and $E^i(k)$ at $331$ms, after $14$ s of free decay, and for a charge-3 persistent current. 
}  
\vspace{-4mm}
\end{figure}

The spectra of Fig.~\ref{fig3} are obtained from various times of the simulation and calculated using spatial grids of $1811^2$ points separated by $\xi/4=0.1\,\mu$m, where the $\xi$ is the healing length. Each curve shows the spectrum of a 2D slice through $z=0$, although the spectra are little changed by averaging slices through the BEC.  The ultraviolet (large $k$) $E^i(k)\propto k^{-3}$ region of the spectrum is a conspicuous feature once vortices are present. This power law is a universal property of a quantized vortex core in a compressible 2D quantum fluid, as analyzed in Ref.~\cite{Bradley2012a}, occurring for $k>k_s\equiv \xi^{-1}$. 
The associated length scale $\sim$$2\pi\xi$ thus serves to distinguish between scales where the system's physical characteristics are dominated by motion of point-like vortices ($k < k_s$), and those where characteristics derive from the structure of individual vortex cores ($k > k_s$).  The ultraviolet power law only plays a role in the energy spectrum through its amplitude, which is proportional to the total vortex number~\cite{Bradley2012a}. The only mechanisms that can appreciably change the incompressible energy for $k>k_s$ are creation and loss of free vortices.  

As stirring injects kinetic energy into the system a Kolmogorov $k^{-5/3}$ power-law spectrum develops in the $k<k_s$ region, and is determined by the vortex configuration~\cite{Bradley2012a}. This spectrum spans a decade in $k$-space and is established by the end of the stir ($\sim$331 ms) which is also when total incompressible kinetic energy is maximal. Logarithmically bilinear spectra are also obtained after ramping off the central barrier (not shown).  Post stir, Fig.~\ref{fig3} (inset)  indicates a slow loss of energy with approximate preservation of the Kolmogorov power law.  Eventually the system spectrally condenses via generation of a metastable  persistent current with three units of circulation.

%============================================================================
%============================================================================

Three additional wavenumbers are indicated in  Fig.~\ref{fig3}.  The cross-sectional radial thickness of the BEC approximately corresponds to the length scale $25\, \mu$m~$=2\pi/k_T$. At high wavenumbers, $k_{\xi}=2\pi/\xi$ corresponds to the scale of the healing length $\xi=0.42\, \mu$m, the approximate size of the smallest features (e.g. vortex cores) supported by a BEC. Finally, a wavenumber $k_F$ is 
associated with the peak in the incompressible spectrum at $\sim 181$ms, as we now describe.

In the classical theory of forced 2D turbulence~\cite{2DCTreviews}, 
spectrally localized forcing is related to the injection rates of enstrophy ($\eta$) and energy ($\epsilon$) density as $k_F=\sqrt{\eta/\epsilon}$~\cite{Les2008.Turbulence}.  In our numerical results, the forcing that precedes the full development of the logarithmically bilinear spectrum is associated with the spectral peak at $181$ ms in Fig.~\ref{fig3}. At $208$ ms the peak has dispersed and the spectrum is already approximately logarithmically bilinear.  We estimate the location of $k_F$ from the numerically computed changes in total incompressible kinetic energy, $\Delta E^i \approx 2.9\times 10^{-3}\,\mu_0\!\cdot\! N$, and enstrophy, $\Delta \Omega \approx 0.963\times 10^{-3}\,\mu_0\!\cdot\! N/\xi^{2}$, occurring during this $27$-ms time interval. We find $k_F\equiv \sqrt{\Delta \Omega/\Delta E^i}=0.57 \xi^{-1} \simeq 2\pi/(11\xi)$, shown in Fig.~\ref{fig3}, and coinciding with the spectral peak. 

The physical mechanism for injection of energy and vorticity into our BECs involves coupling between pairs of opposite-circulation vortices and acoustic energy.  In the stirring process, density waves are first generated in the BEC, most prominently where the fluid density approaches zero.  These density waves decay to vortices (see Movie S1 \cite{SM}) in two characteristic ways.  First, a density wave can develop into a localized dark soliton and then decay to a vortex dipole.  A second injection mechanism involves the decay of a density wave near the central barrier into a single vortex within the fluid and a partner antivortex pinned by the barrier.

Empirically we find that the length scale $\sim$2$\pi/k_F$ coincides with the separation of phase singularities created from the decay of a localized sound pulse into a vortex dipole, visible in Movie S1 \cite{SM}.  Examining the instances of vortex dipole creation from sound during the stir period, we find dipole lengths $d$ in the range 6.7$\xi$ to 11$\xi$, suggesting an injection of incompressible energy near a wavenumber $k\sim 2\pi/d$. During the stir, there is one case of a vortex dipole annihilating irreversibly to sound at $t=0.23$ s, where the dipole length is  $d \sim6.7\xi$.  Two transient events during the nominal constant enstrophy period discussed below correspond to dipole annihilation at $\sim6.7\xi$ and $\sim8.9\xi$. Furthermore, the superfluid density modulations preceding vortex nucleation at the 181-ms spectral peak have a length scale of approximately $11\xi \simeq 2\pi/k_F$; see Fig.~S2 \cite{SM}.  Taken together, these observations indicate that forcing involves efficient energy and enstrophy transfer from the compressible to the incompressible fluid components for wavenumbers $k_F\lesssim k\lesssim k_s$.

\begin{figure}[t]
\begin{center}
\includegraphics[width=.48\textwidth]{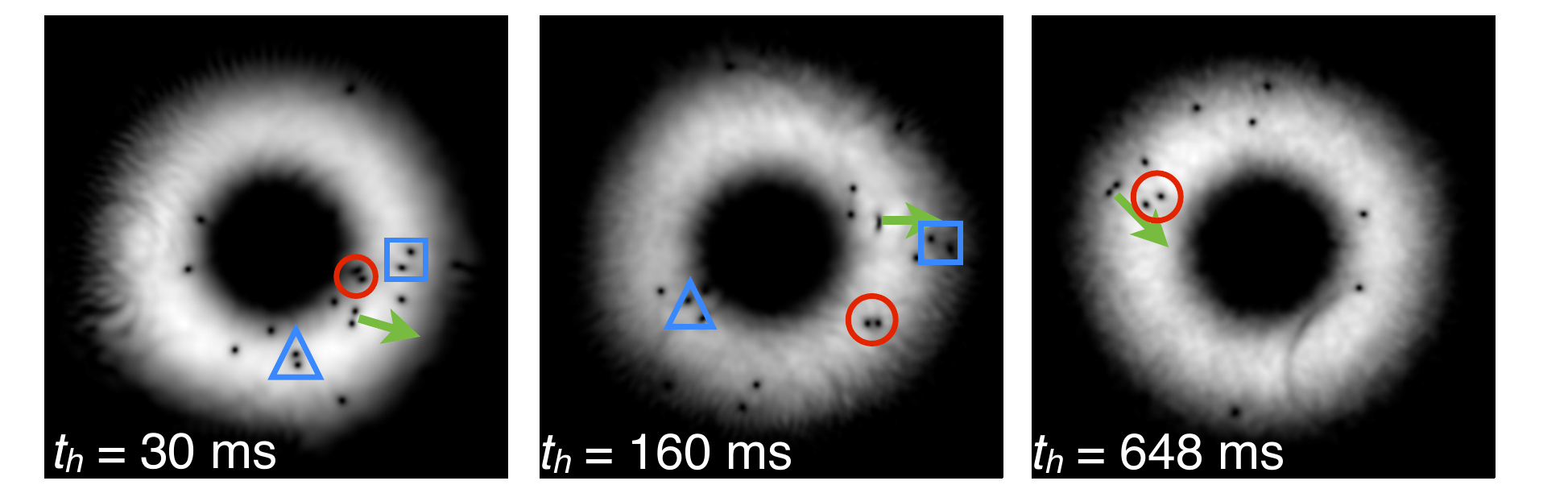}
\caption{\label{fig5} 
Numerically obtained BEC column density shown for three hold times.  In each 96-$\mu$m-square image, symbols indicate clusters of same-sign vortices, either co-rotating (red), or counter rotating (blue) with the stir. Each shape represents the same cluster at different times. Vortex dipoles and their propagation directions are indicated by green arrows. At $t_h=648$ ms, one cluster remains after having traveled clockwise halfway around the BEC.  See Fig.~S3 \cite{SM}.
}
\end{center}
\vspace{-7mm}
\end{figure}

The conservation of enstrophy in 2D turbulence corresponds to conservation of vortex number in 2DQT \cite{Num2010.JLTP158.415,Bradley2012a}.  Between $\sim$300 ms and $\sim$600 ms after the stir begins, the total vortex number is nominally constant in our simulation. When vortex-antivortex annihilation occurs, the resulting sound pulses are quickly (within $\sim$20 ms) refocused by the inhomogeneous density, regenerating the vortices.  We thus identify this 300-ms period of nominally constant vortex number with enstrophy conservation.  During this period, we numerically observe four instances of the formation of two-vortex clusters (same-sign vortices).   Fig.~\ref{fig5} shows three two-vortex clusters and a vortex dipole (opposite-sign vortices) present 30 ms after the end of stir.  This dipole exists for 20 ms, while the longest-lived of the two-vortex clusters exists for 630 ms. The vortices of this cluster orbit each other 15 times, travel together halfway around the BEC, and eventually dissociate upon colliding with a vortex dipole; see Movie S1 \cite{SM} and Fig.~\ref{fig5}.  We see in Fig.~\ref{fig2}(a) and (b) that there are large regions free of vortices, and regions where many vortices are densely packed, which is indicative of clustering, but we are not able to measure the vortex circulation directions to directly confirm this. Clustering was, however, experimentally observed in Ref.~\cite{Nee2010.PRL104.160401} in the form of dipolar clusters reproducibly generated by a moving obstacle. In the present case of circular forcing, numerically we observe long-lived clusters emerging intermittently within otherwise irregular flows, in a manner suggestive of Onsager's statistical hydrodynamics result \cite{Ons1949.NC6s2.279}.

Over the tens of seconds after stirring stops, free vortices decay either by being pinned to the obstacle beam, damping at the outer BEC boundary, or annihilating with free or pinned vortices of opposite sign. This results in the formation of a persistent current, reflecting the net angular momentum imparted by the stirring. The development of this superflow represents the growth of large-scale flow originating from small-scale forcing in an annular geometry, and serves as a check of our numerical procedures and interpretations. The mean number of vortices (pinned and free) for $t_h = 23$ s is 3.5 in the experiment, and 5 in the simulation. For $t_h = 43$ s, these values decline to 2.5 and 3 respectively.

%============================================================================
%============================================================================

Previous investigations of 2DQT have centered on numerically obtaining kinetic energy spectra, but these have been inconclusive regarding the possibility of an inverse energy cascade in 2DQT, conservation of enstrophy, and correspondence between spectra and vortex dynamics.   In our study, simulation of the experimentally realized forcing shows the development of an inertial range.  Additionally, small-scale forcing within the trap enables the clustering of vortices; as clustering suppresses vortex-antivortex annihilation, it provides a mechanism to enable enstrophy conservation in a compressible superfluid.  Regarding the possibility of a compressible superfluid supporting an inverse energy cascade, energy flux calculations provide the most direct route to analyzing cascades, although such an approach for a trapped BEC with an inhomogeneous density distribution is an open problem \cite{flux}.    However, the following observations are consistent with the existence of an inverse energy cascade in our system near the end of the stir: (i) vortex dipole recombination is suppressed and thus there is little dissipation over a forcing range $k_F$ to $k_s$; (ii) $E^i(k)\propto k^{-3}$ for $k>k_s$, a range that cannot support incompressible energy flux \cite{Bradley2012a}; (iii) enstrophy is nominally conserved, and kinetic energy spectral developments occur primarily for $k<k_s$; and (iv) $E^i(k)\propto k^{-5/3}$ for $k<k_s$, a signature of conserved energy transfer across the associated scales.  
%============================================================================
\addtocounter{figure}{-4}
\def\figurename{Figure S$\!\!$}

\begin{figure*}[!htb]
\begin{center}
\begin{large}
\textbf{Supplemental Material:  Characteristics of Two-Dimensional Quantum Turbulence in a Compressible Superfluid}\\[2mm] 
\end{large}
T.~W.~Neely$^1$, A.~S.~Bradley$^2$, E.~C.~Samson$^1$, S.~J.~Rooney$^2$, E.~M.~Wright$^1$, K.~J.~H.~Law$^3$,\\
R.~Carretero-Gonz\'{a}lez$^4$, P.~G.~Kevrekidis$^5$, M.~J.~Davis$^6$, B.~P.~Anderson$^1$\\[2mm]
\begin{small}
{\em $^1$College of Optical Sciences, University of Arizona, Tucson, AZ 85721, USA\\
$^2$Jack Dodd Centre for Quantum Technology, Department of Physics, University of Otago, Dunedin 9016, New Zealand\\
$^3$Mathematics Institute, University of Warwick, Coventry CV4 7AL, UK\\
$^4$Department of Mathematics and Statistics, San Diego State University, San Diego, CA 92182, USA\\
$^5$Department of Mathematics and Statistics, University of Massachusetts, Amherst, MA 01003, USA\\
$^6$School of Mathematics and Physics, University of Queensland, Qld 4072, Australia}\\[2mm]
Here we provide material supplementing the main text [access information to be provided by publisher]. The\\
supplemental material consists of additional description of numerical methods, three figures, and one movie.
\end{small}
\end{center}
\end{figure*}
%============================================================================
Our observations indicate that characteristics of forced and decaying 2DQT in compressible quantum fluids may be analogous to those of 2D turbulence in incompressible classical fluids.  In particular, growth of large-scale flow, aggregation of vorticity, nominal enstrophy conservation, and an energy spectrum with $k^{-5/3}$ and $k^{-3}$ spectral features occur with suitable forcing. The vortex clusters are suggestive of Onsager's predictions, indicating a new link between Onsager's analysis of 2D point-vortex dynamics \cite{Ons1949.NC6s2.279} and the theory of 2D turbulence initiated by Kraichnan, Leith, and Batchelor \cite{2DT}. Our observations motivate further investigations of 2DQT, with future work focusing on energy fluxes, the roles of dissipation and boundary conditions, and direct experimental observations of turbulent vortex dynamics of a 2D quantum fluid.

\acknowledgements

We thank Colm Connaughton and Sergey Nazarenko for useful discussions, and the US National Science Foundation grants PHY-0855677 and DMS-0806762, the US Army Research Office, The Marsden Fund, The Royal Society of New Zealand, and The New Zealand Foundation for Research, Science, and Technology contracts UOOX0801 and NERF-UOOX0703 for funding.\\
\\
References follow Supplemental Material.
\newpage
%%%%%%%%%%%%%%%%%%%%%%%%%%%%%%%

\textbf{Theoretical background}.  The damped GPE theory used in our simulations can be derived rigorously for the dilute-gas BEC through a detailed treatment of reservoir interactions within the Wigner phase-space representation \cite{refS31}, by neglecting thermal noise. An approximate and practical stochastic Gross-Piteavskii theory can be obtained~\cite{refS32}
by: (i) neglecting the particle-conserving reservoir interaction processes (\emph{scattering} terms) that are known to be small in the quasi-equilibrium regime; and (ii) neglecting the weak spatial and time dependence of the damping parameter. This allows the damping parameter to be calculated \emph{a priori}, once the reservoir parameters are known. The resulting stochastic Gross-Pitaevskii equation (SGPE) has been used to study spontaneous vortex formation \cite{refS33} during Bose condensation, and vortex dynamics [21, 26] at high temperature.  The SPGPE is derived by treating all atoms above an appropriately chosen energy cutoff $\ecut$ as thermalized (incoherent region) with temperature $T$ and chemical potential $\mu$, leading to a grand-canonical description of the atoms below $\ecut$ (coherent region).  A dimensionless  rate $\gamma\equiv \gamma(T,\mu,\ecut)$ describes Bose-enhanced collisions between thermal reservoir atoms and atoms in the BEC.  

However, it is not known how to extract a well-defined condensate orbital from the SGPE trajectories in high-temperature systems containing vortices. To extract physical information about the condensate dynamics we nelgelct the thermal noise to obtain the damped Gross-Piteavskii equation of motion. We thus arrive at a tractable, microscopically determined equation for the condensate orbital, with \emph{a priori} determined reservoir parameters. 
After a trivial shift of energy by the chemical potential $\mu$ we obtain the equation of motion for the wavefunction $\psi(\textbf{r},t)$
\begin{equation}\label{eom}
i\hbar\frac{\partial \psi(\brr,t)}{\partial t}=(i\gamma-1)(\mu-L)\psi(\brr,t),
\end{equation} 
where the operator $L$
\begin{equation}
L\psi(\brr,t)\equiv \left(-\frac{\hbar^2}{2m}\nabla^2+V(\brr)+g|\psi(\brr,t)|^2\right)\psi(\brr,t),
\end{equation}
 is the generator of GPE evolution for atoms of mass $m$ in an external potential $V(\brr)$.
The interaction parameter is $g=4\pi\hbar^2 a/m$, for s-wave scattering length $a$.  
In most cases the damping parameter is small ($\gamma\ll 1$), and is typically much smaller than any other timescales characterizing the  evolution.   

In applying our approach to modeling the experiment of the main text, we find the self-consistent parameters $\mu=34\hbar\bar{\omega}$, $\epsilon_{\rm cut}=83\hbar\bar{\omega}$, for geometric mean $\bar{\omega}=(\omega_r^2\omega_z)^{1/3}$, to describe a system of $N=2.6\times 10^6$ atoms held at temperature $T=0.9T_c$ in the combined trap, giving the damping parameter $\gamma=7.96 \times 10^{-4}$. The potential is characterized by half-width $\sigma_0=16.3\,\mu$m, and is well contained within the coherent region: $\sigma_0\ll R_{\rm cut}=\sqrt{2\epsilon_{\rm cut}/m\omega_r^2}=73\,\mu$m, which is the spatial cutoff imposed by $\epsilon_{\rm cut}$. It thus has no other significant effect on the incoherent region. These parameters give an initial state containing $\sim 6\times 10^5$ coherent region atoms, with the remainder in the incoherent region. 

There are some technical limitations of this approach.  First, the number of atoms in the simulations is approximately constant due to the fixed chemical potential of the thermal reservoir, while the BEC number in the experiment first grows as $T/T_c$ is reduced to $\sim0.6$, then decays with a 1/$e$ lifetime of 24(3) s. A related issue is the slow spatial drift of the barrier beam of the experiment, which can decrease the number of vortices that can be stably pinned to the barrier, providing a loss mechanism for the persistent current at long times; this latter aspect of our work will be discussed in a separate publication.  However, our procedure is suitable for simulating the BEC conditions early in the stir and hold process, and we do not expect detailed modeling of the evaporative cooling stage to alter our main results.\\

\textbf{Movie S1:  Dynamics of a forced, damped BEC.}   A movie of the damped Gross-Pitaevskii equation dynamics may be viewed online at [URL to be provided by Physical Review].  The movie shows the column density and phase profile in the $z=0$ plane generated for a simulation of damped Gross-Pitaevskii equation for the experimental parameters. A number of events are visible, notably:
\begin{enumerate}[i)] 
\item $t=140$ ms: The first evidence of a boundary vortex appears at the {\em inside} boundary of the BEC.  This is most apparent in the phase profile.
\item $t=160$ ms: A prominent density modulation feature develops.
\item $t=166$ ms: This time is half-way through stir sequence.
\item $t=190-210$ ms: The first vortices injected into the BEC via sound decay are visible.
\item $t=220$ ms: Many vortices are now found in the bulk superfluid. 
\item $t=230$ ms: Vortex dipole annihilation.
\item $t=260$ ms: Two long-lived vortex aggregates have formed ($\sim$3 o'clock and 4 o'clock positions).

\item $t=390-410$ ms: A vortex approaches the inner BEC boundary ($\sim$4 o'clock position) and is captured by the central barrier, increasing the circulation pinned to the barrier by one unit.   In the process, a sound pulse is released into the BEC.
\item $t=420-450$ ms: A vortex dipole forms as two vortices of opposite circulation approach each other ($\sim$ 7 o'clock position), then undergoes self-annihilation followed by revival at the outer BEC boundary.
\item $t=480-520$ ms: A vortex dipole forms ($\sim$ 3 o'clock position), then undergoes self-annihilation followed by revival at the outer BEC boundary.
\item $t=590-640$ ms: A vortex dipole forms ($\sim$ 8 o'clock position), then undergoes self-annihilation followed by revival at the outer BEC boundary.
\item $t=600$ ms: Thermalization of the sound field across the whole system has occurred by this time.
\item $t=790$ ms: Vortex dipole annihilation.
\item $t=1000$ ms: A two-vortex cluster, a dipole, and a free vortex collide. Vortex exchange occurs; the collision results in the same vortex structures emerging. 
\item $t=1.980$ s: A vortex collision results in a vortex being slightly tilted with respect to the $z$ axis ($\sim$ 4 o'clock position), showing that this system is not strictly two-dimensional.  Note that this vortex returns to an orientation along the $z$ axis within about 50 ms.\\[5mm]
\end{enumerate}

%============================================================================
\begin{figure}[!t]
\includegraphics[width=.84\columnwidth]{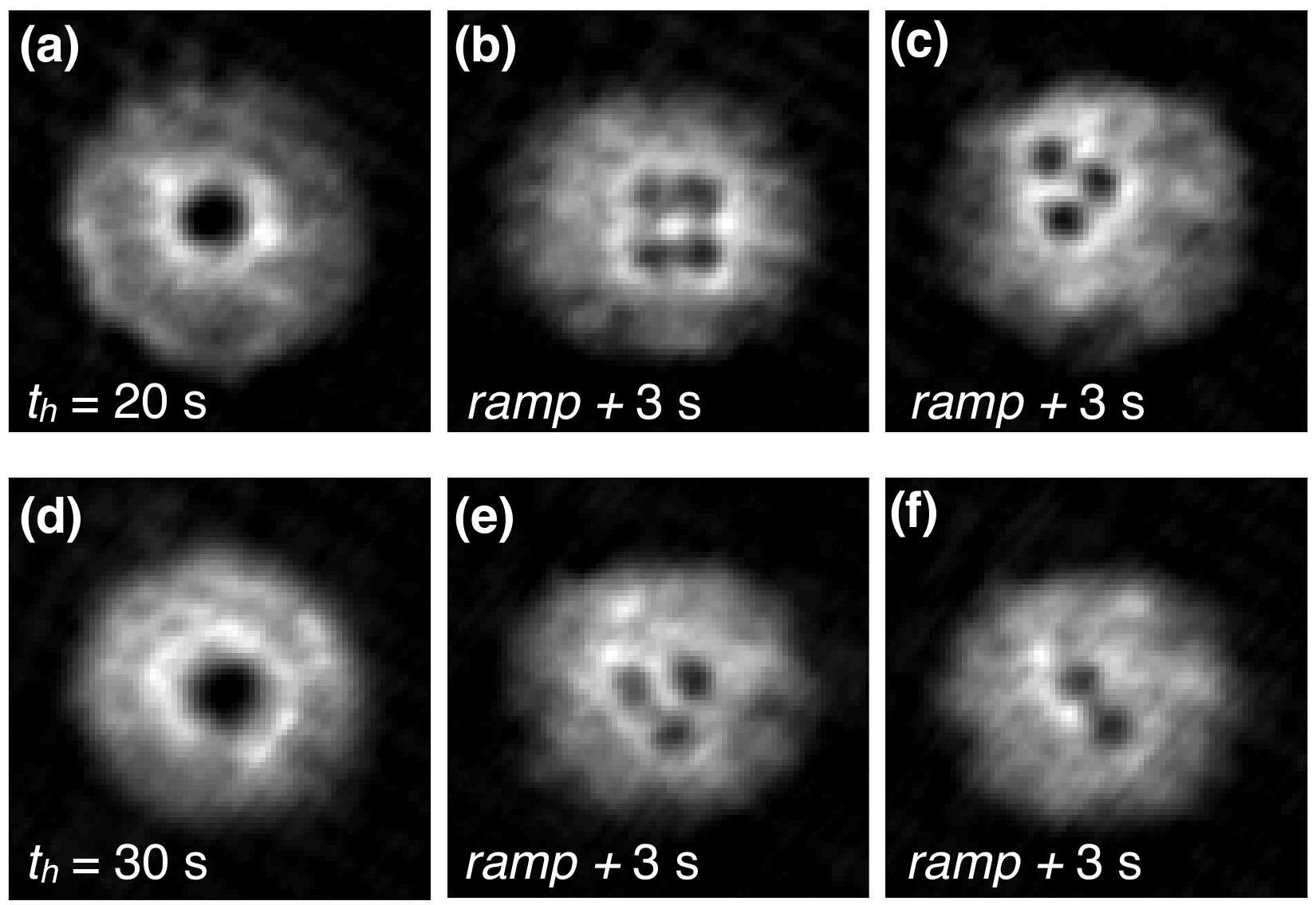}
\caption{\label{figS1} 
Vortices observed in experimental data.  Here, BECs are held in the annular trap for 20 s prior to the central barrier ramp down, trap release, expansion, and imaging. In (a) the expansion and imaging procedure takes place immediately after the barrier ramp.  The dark region in the center of the BEC corresponds to multiple units of circulation, and is not due to the presence of the barrier itself; i.e., with no circulation about the barrier, a hole would not appear in the BEC after ballistic expansion. In (b) and (c), BECs are held for an additional 3 s in the trap before the expansion imaging procedure.  This additional hold time allows the vortices that comprise the region of vorticity shown in (a) to dissociate and become experimentally observable.  (d)-(f) show similar images for $t_h = 30$ s.  
}
\vspace{3mm}
\end{figure}
%============================================================================

%============================================================================
\begin{figure}[!b]
\includegraphics[width=.5\columnwidth]{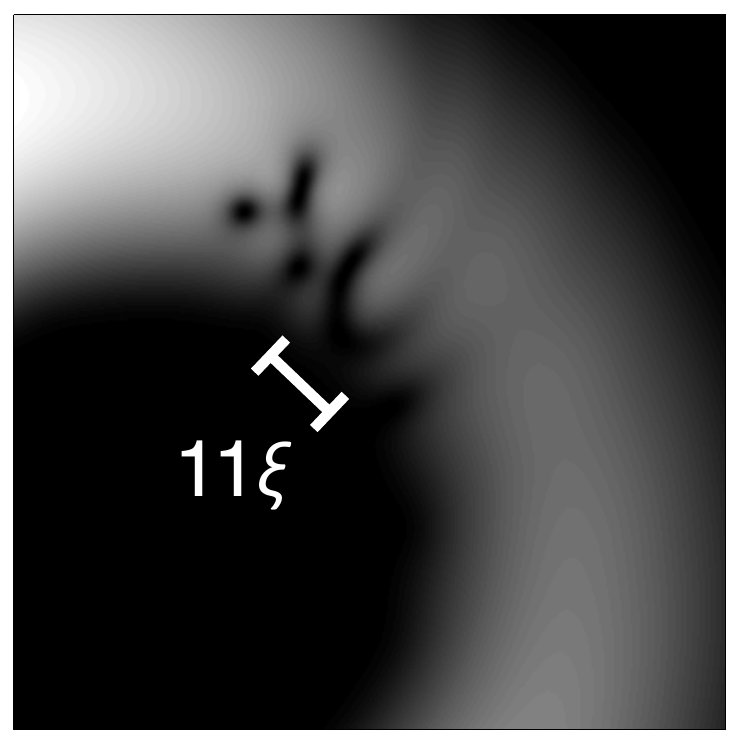}
\caption{\label{figS3} 
Numerically obtained column density $181$ ms after stirring begins. The bar indicates the scale of forcing calculated from the energy and enstrophy injected during the stir (see main text). 
40-$\mu$m-square field of view.  %%%% scale bar given, field of view doesn't need to be stated
}
\end{figure}
%============================================================================

%============================================================================
\begin{figure}[!t]
\vspace{-0mm}
\begin{center}
\includegraphics[width=.48\textwidth]{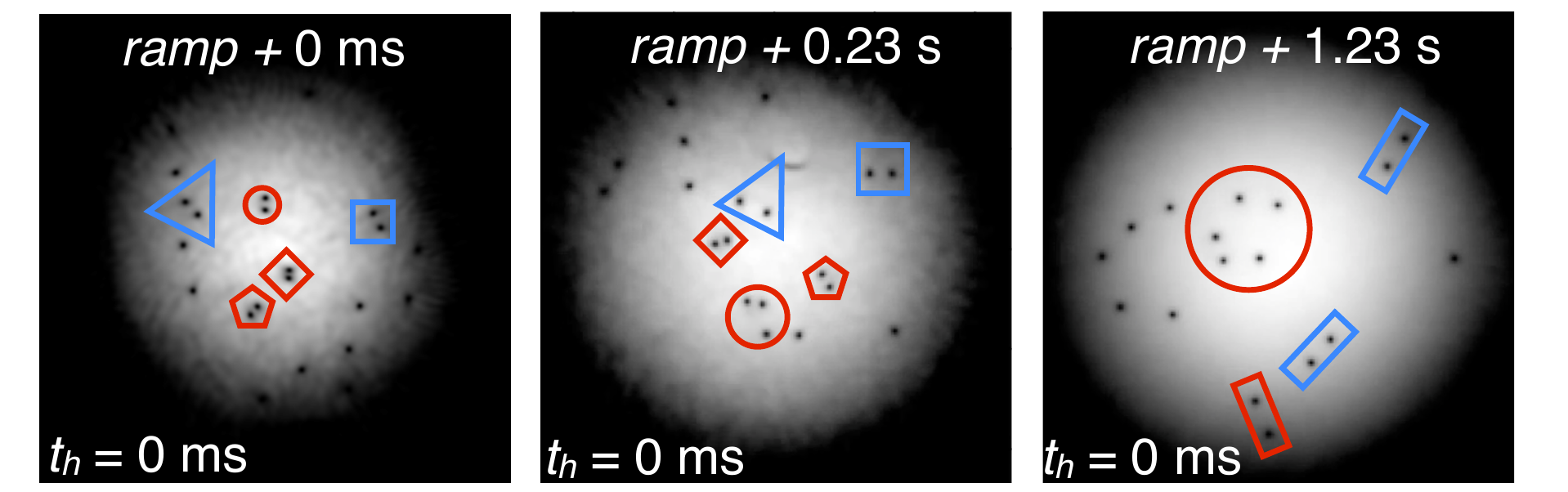}
\caption{\label{figS4} 
Vortex clusters seen for a main sequence hold time of $t_h= 0$ ms, followed by the 250-ms barrier ramp-down, and an additional hold time of 0 ms, 230 ms, and 1.23 s in the harmonic trap (96-$\mu$m-square images).  In the left-most image, five two-vortex clusters are seen. Only two or three of these clusters existed at $t_h=0$ ms  \emph{prior} to the beam ramp (red circle, blue triangle and square of Fig.~4 of the main text), while the remainder formed from vortices that were pinned to the beam and then released into the quantum fluid.  By 230 ms after the laser barrier has been ramped completely off, one of the two-vortex clusters has acquired a third vortex (red circle of middle image), and these three vortices orbit each other. This structure eventually grows to form a loose five-vortex cluster shown  1.23 s after the barrier ramp.   This cluster persists for about 200 ms.  The remaining clusters have dissociated by this time, but three new two-vortex clusters have formed (rectangles).   
}
\end{center}
\end{figure}
%============================================================================

\textbf{Experimental winding number determination}.  Multi-quantum vortices are energetically unstable, thus loose clustering is favorable to perfect co-location of multiple vortices.  We make use of this energetic instability in order to determine the size of the persistent current formed and its subsequent time evolution in the experiment, as shown in Figure S1.    By enabling extra hold time between beam ramp down and expansion, we can count the total number of free and pinned vortices for any hold time.  For long-enough hold times where the number of free vortices has dropped to much less than one per image (on average), the observed vortices can be attributed to current pinned at the barrier, particularly if they appear clustered about the position of the barrier as shown in Fig.~S1.  We note that observations of such regular structures of vortices, as shown, do not always occur after beam ramp down, and the vortex distribution is often more irregular.  As stated in the main text, the mean number of vortices experimentally observed at $t_h = 23$ s is 3.5.  If instead we remove the central barrier at the {\em beginning} of the hold period, and let the system evolve in a purely harmonic trap for 23 s, the mean numbers of vortices observed becomes 1.2.  The fluid circulation is thus maintained at significantly higher levels in the annular trap, justifying a description of this state as a persistent current.\\[0mm]

\textbf{Density modulations during forcing}.  Figure S2 shows the stir-induced density modulations observed in the numerical simulations.  These density modulations decay to vortices.  As density modulations at a wavenumber $k$ correspond to {\em compressible} energy at that wavenumber, we expect an influx of energy into the incompressible regime at a wavenumber that corresponds to these modulations, as discussed in the main text.\\[-0mm]

\textbf{Vortex clusters and dynamics}.
In Fig.~3 of the main text we show images from our simulation that contain vortex clusters that occur after the completion of the stir in the annular trap.  Alternatively, we may examine the distribution of vortex clusters in the gas after the ramp down of the stirring potential to determine whether such clusters might occur in the experimental system subsequent to beam ramp down.  Clusters do indeed appear in such numerical data, as Fig.~S3
shows, and are not limited to just two vortices per cluster.

%%%%%%%%%%%%%%%%%%%%%%%%%%%%%%%


\begin{thebibliography}{10}

\bibitem{Les2008.Turbulence}
M. Lesieur, {\em {Turbulence in Fluids}} (Kluwer Academic Publishers,
  Netherlands, 4th ed., 1990).

\bibitem{2DT}
R. Kraichnan, Phys. Fluids {\bf 10},  1417  (1967);
C. Leith, Phys. Fluids {\bf 11},  671  (1968);
G. Batchelor, Phys. Fluids {\bf 12},  II  (1969);
R. Kraichnan and D. Montgomery, Rep. Prog. Phys. {\bf 43},  547  (1980).

\bibitem{2DCTreviews}
J. Sommeria, Les Houches Summer School 2001: New Trends in Turbulence {\bf 74},
385  (2001);  P. Tabeling, Phys. Rep. {\bf 362},  1  (2002); G. Boffetta and R. E. Ecke, Ann. Rev. Fluid Mech. {\bf 44}, 427 (2012).

\bibitem{Kra2008.PF20.056602}
W. Kramer, H. Clercx, and G. van Heijst, Phys. Fluids {\bf 20},  056602
  (2008).

\bibitem{Mar1998.PRL80.3964}
B. Martin {\em et al.}, Phys. Rev. Lett. {\bf 80},  3964
 (1998).

\bibitem{Riv2003.PRL90.104502}
M. Rivera {\em et al.}, Phys. Rev. Lett. {\bf 90},  104502
  (2003).
  
\bibitem{Sha2005.PRE71.046409}
M. Shats, H. Xia, and H. Punzmann, Phys. Rev. E {\bf 71},  046409  (2005).

\bibitem{Som1988.Nat331.689}
J. Sommeria, S.~D. Meyers, and H.~L. Swinney, Nature {\bf 331},  689  (1988).

\bibitem{Mil1992.PRA45.2328}
J. Miller, B. Weichman, and M.~C. Cross, Phys. Rev. A {\bf 45},  2328  (1992).

\bibitem{Bar2001.book.Superfluid}
C.~F. Barenghi, R.~J. Donnelly, and W.~F. Vinen, eds., {\em {Quantized vortex
  dynamics and superfluid turbulence}} (Springer, Berlin, New York,
  2001).

\bibitem{Vin2002.JLTP128.167}
W.~F. Vinen and J.~J. Niemela, J. Low Temp. Phys. {\bf 128},  167  (2002).

\bibitem{enstrophy}
In two-dimensions, enstrophy $\Omega = \int dx\, dy\, \omega^2$ is a measure of
  the vorticity $\omega = [ \nabla \times v ]_z$ about the $z$ axis, where $v$
  is the fluid velocity in the plane normal to $z$. 
  
\bibitem{Ons1949.NC6s2.279}
L. Onsager, Il Nuovo Cimento {\bf 6 suppl 2},  279  (1949).

\bibitem{Eyi2006.RMP78.87}
G. Eyink and K. Sreenivasan, Rev. Mod. Phys. {\bf 78},  87  (2006).

\bibitem{Naz2006.JLTP146.31}
S. Nazarenko and M. Onorato, J. Low Temp. Phys. {\bf 146},  31  (2006);

\bibitem{Hor2009.PRA80.023618}
T.-L. Horng {\it et~al.}, Phys. Rev. A {\bf 80},  023618  (2009).

\bibitem{Num2010.JLTP158.415}
R. Numasato and M. Tsubota, J. Low Temp. Phys. {\bf 158},  415  (2010).

\bibitem{Num2010.PRA81.063630}
R. Numasato, M. Tsubota, and V. L'vov, Phys. Rev. A {\bf 81},  063630  (2010).

\bibitem{Gasenzer}
B. Nowak, J. Schole, D. Sexty, T. Gasenzer, arXiv:1111.6127 (2011).


\bibitem{Nee2010.PRL104.160401}
T.~W. Neely {\it et~al.}, Phys. Rev. Lett. {\bf 104},  160401  (2010).

\bibitem{Roo.PRA84.023637}
 S.~J. Rooney {\em et al.}, Phys. Rev. A
  {\bf 84},    (2011).


\bibitem{KetterleQT}
C. Raman {\em et al.}, Phys. Rev. Lett. {\bf 87}, 210402 (2001).

\bibitem{Hen2009.PRL103.045301}
E. Henn {\it et~al.}, Phys. Rev. Lett. {\bf 103},  045301  (2009).



\bibitem{Bla2008.AIP57.363}
P.~B. Blakie {\it et~al.}, Adv. in Phys. {\bf 57},  363  (2008).


\bibitem{Pit1958}
L.~P. Pitaevskii, Zh. Eksp. Teor. Fiz. {\bf 35},  408  (1958).


\bibitem{Roo.PRA81.023630}
S.~J. Rooney, A.~S. Bradley, and P.~B. Blakie, Phys. Rev. A {\bf 81},  023630
  (2010).
  
\bibitem{SM}
See Supplemental Material at [URL will be inserted by publisher] for additional information.


\bibitem{Nor1997.PF9.2644}
C. Nore, M. Abid, and M.~E. Brachet, Phys. Fluids {\bf 9},  2644  (1997).


\bibitem{Bradley2012a}
A. S. Bradley and B. P. Anderson, arXiv:1204.1103 (2012).


\bibitem{flux} See Ref.~\cite{Num2010.PRA81.063630} for one approach to calculating energy flux in a \emph{homogeneous} BEC evolving under decaying 2DQT.\\

\begin{center}{\textbf{Supplemental Material References}}\end{center}


\bibitem{refS31}
C.~W. Gardiner and M.~J. Davis, J Phys B {\bf 36},  4731  (2003).

\bibitem{refS32}
A.~S. Bradley, C.~W. Gardiner, and M.~J. Davis, Phys. Rev. A {\bf 77},  033616
  (2008).

\bibitem{refS33}
C.~N. Weiler {\it et~al.}, Nature {\bf 455},  948  (2008).


\end{thebibliography}
\end{document}